# An improved design method for conventional straight dipole magnets*

Ying-shun Zhu(朱应顺) [1;1)], Wen Kang(康文) [1,2], Fu-san Chen(陈福三) [1], Wan Chen(陈宛) [1], Xi Wu(吴锡) [1,2], Mei Yang(杨梅) [1,2]

[1] Key Laboratory of Particle Acceleration Physics and Technology，Institute of High Energy Physics, Chinese Academy of Sciences, Beijing 100049, China

[2] China Spallation Neutron Source, Institute of High Energy Physics, Chinese Academy of Sciences, Dongguan 523803, China

**Abstract:** The standard design method for conventional straight dipole magnets is improved in this paper. The good field region is not symmetric with respect to the magnet mechanical center, and its width is not enlarged to include the beam sagitta. The integrated field quality is obtained by integrating the field along nominal beam paths. 2D and 3D design procedures of the improved design method are introduced, and two application examples of straight dipole magnets are presented. It is shown that the differences in integrated field quality between different field integration paths cannot be neglected. Compared with the traditional design method of straight dipole magnets, the advantage of the improved method is that the integrated field quality is accurate; the pole width, magnet dimension and weight of a straight dipole magnet can be reduced.

## 1 Introduction

Dipole magnet is one of the most fundamental and commonly used magnet type in high energy accelerators [1-3]. A dipole magnet produces uniform magnetic field to bend the beam, so it is also referred as bending magnet. Unlike other magnet types such as quadrupole magnet, sextupole magnet which are usually straight in the longitudinal direction, a dipole magnet can be either straight or curved.

Compared with the curved dipole magnet, straight dipole magnet is easier to be manufactured and assembled, and higher mechanical precision can be achieved [1, 4]. Thus it is preferred in many applications. So far, a number of conventional dipole magnets (including gradient dipole magnet) in high energy accelerators have been designed and manufactured as straight [4-11].

When a straight dipole magnet is installed in the tunnel, usually the central beam path is offset from the nominal magnet centerline by half the beam sagitta. However, in the standard magnetic design method of straight dipole magnets, the good field region (GFR) is symmetric to the magnet mechanical center both in 2D and 3D. Relative to the one required by the beam optics, the GFR is enlarged to include the beam sagitta. Furthermore, the magnetic field is usually integrated along straight lines longitudinally to obtain the integrated field in the field simulation and field measurement, which is different from the method used in curved dipole magnets [12-14]. The integrated field uniformity obtained along straight lines is not the actual field quality of a straight dipole magnet because the real beam path is curved. So the difference in the integrated field quality between the straight line integration and curved line integration is ignored in the traditional design method of straight dipole magnets.

* Supported by China Spallation Neutron Source and Key Laboratory of Particle Acceleration Physics and Technology, Institute of High Energy Physics, Chinese Academy of Sciences.
1) E-mail: yszhu@ihep.ac.cn

An existing improvement in the design of straight dipole magnet is to divide a straight magnet into several shorter straight sections with smaller bending angle and smaller beam sagitta in each section [15-16]. This can avoid the large beam sagitta and larger magnet size in single straight magnet. However, this method increases the magnet sections and complicates the magnet manufacture and alignment process. In addition, the beam sagitta is still included in the good field region, and the field is also integrated along straight lines to get the integrated field in this method.

The main purpose of this paper is to improve the current design method of conventional straight dipole magnets. The relationship of beam paths with respect to the straight magnet geometry is carefully considered. The good field region is not symmetric to the magnet mechanical center and is not enlarged to include the beam sagitta. The integrated field is obtained by integrating the magnetic field along ideal curved beam paths. The difference in integrated field quality between the improved and traditional design methods in the same straight dipole magnet is investigated.

**2 General description of the improved design method**

The width of GFR is an important parameter in the design of a dipole magnet because it is directly related to the pole width and magnet transverse size. In the traditional design method of straight dipole magnets, the width of total GFR is the sum of beam sagitta and the real GFR required by the beam optics [4-11, 15-16]; in addition, the center of GFR is consistent with the magnet mechanical center.

The basic idea of the improved design method is that the relationship of curved beam paths with respect to the straight magnet geometry is considered, and only the GFR required by the beam optics is used. The center of GFR does not coincide with the magnet mechanical center both in 2D and 3D. The sketch map of the curved beam trajectory passing through a straight dipole magnet is shown in Fig. 1. The apex of the central beam path (Point *A*) is offset from the nominal magnet centerline by half the beam sagitta, which is the same as that in the traditional design method [5].

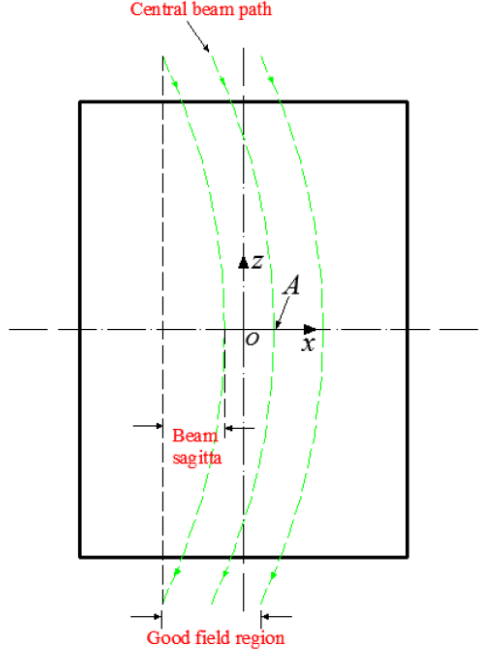

Fig. 1. Beam trajectory passing through a straight dipole magnet

It can be seen in Fig. 1 that the transverse relationship of the required GFR with respect to the straight magnet geometry varies at different longitudinal positions, so 2D transverse field uniformity in GFR also differs. It is convenient to correspond the middle plane of the magnet in longitudinal direction ($z$=0) to the plane in 2D design. Then 2D magnetic design and field analysis can be performed similar to the process in the traditional design method.

Denoting the value of beam sagitta as $s$, the width of required GFR by the beam optics as $w$, and assuming that the geometric centerline of the magnet correspond to $x$=0 in the transverse direction, then the range of GFR in 2D field analysis is from $x_1=s/2-w/2$ to $x_2=s/2+w/2$. It should be noted that the center of GFR is not consistent with the magnet mechanical center in the improved design method. For comparison, the range of enlarged GFR in the traditional design method is from $x'_1=-s/2-w/2$ to $x'_2=s/2+w/2$ in 2D field analysis.

The 3D magnetic modeling of straight dipole magnet in the improved design method is similar to the one in the traditional design method. But in the field analysis, the integral field should be obtained by integrating the magnetic field along curved beam paths, and the transverse offset range of the magnetic field integration path is $w$. For comparison, the field is integrated along straight lines longitudinally in the traditional design method, in which the transverse offset range of the field integration path is equal to $w+s$. Thus the field analysis process in the improved design method is quite similar to the one in a curved dipole magnet. Pole end chamfer is still necessary to improve the integrated field uniformity and meet the field quality requirement.

## 3 Application examples
### 3.1 Development of CSNS LR-BB dipole magnets

As an example, the improved design method is applied to develop LR-BB dipole magnets in the China Spallation Neutron Source (CSNS), which is now under construction in China [17]. The LR-BB dipole magnets transport the beam from linac to Rapid Cycling Synchrotron (RCS), and the design requirements are listed in Table 1.

Table 1. Design requirements of LR-BB dipole magnets

| Item | Unit | Value |
| --- | --- | --- |
| Magnet quantity | | 2 |
| Magnet type | | H-type, DC |
| Bending angle | Degree | 25 |
| Magnetic length | m | 1.5 |
| Central field | T | 0.384 |
| Pole gap | mm | 60 |
| Width of good field region | mm | 65 |
| Integrated field uniformity | | ±0.1% |

The calculated beam sagitta of LR-BB dipole magnet is as large as 81.5 mm. It was decided to choose the straight type dipole magnet and use the improved design method in which the horizontal GFR has a width of 65 mm instead of 146.5 mm in the traditional design method.

Soft iron is used for the core of LR-BB dipole magnet. The excitation coil for each pole is made of two pancake-type coils with a total of four layers, and every pancake is cooled by one water circuit. The coils are wound from 13 mm square OFHC copper conductor with a 7 mm diameter water-cooling channel, and the operation current is 334 A.

Since the magnet central field is modest, a pole root narrower than the magnet pole tip is used. The magnetic field optimization is performed using the software OPERA from Cobham Technical Services [18]. As described in Section 2, the range of GFR is from $x_1$=8.25 mm to $x_2$= 73.25 mm in 2D field analysis. Fig. 2 shows the 2D flux lines.

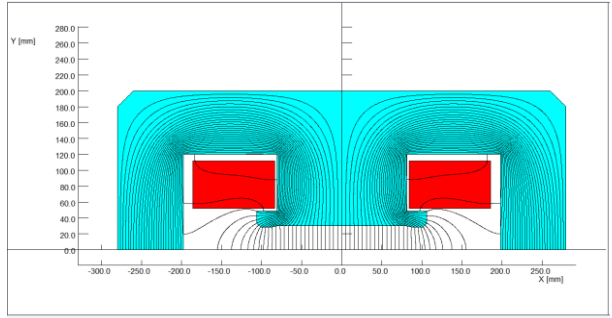

Fig. 2. 2D magnetic flux lines of LR-BB dipole magnet (one half)

The magnetic field homogeneity in the GFR along the transverse axis is calculated. Since the GFR in the improved design method is not symmetric with respect to the magnet mechanical center, the field is higher on the right side than on the left side. However, the peak-to-peak field uniformity of ±0.1% is achieved in 2D simulation. After optimization, the width of pole tip is 202 mm, while it will be 226 mm using the traditional design method. Thus, 24 mm pole width has been saved using the improved design method. The width of LR-BB dipole magnet is 560 mm, and the reduction in the magnet width is about 50 mm.

In 3D magnetic field simulation, the field is integrated along the ideal beam paths with a constant radius of 3437.2 mm. Pole end chamfer is optimized and determined by the field simulation to meet the integrated field quality requirement and is then directly machined on a CNC machine, which is similar to that in a curved dipole magnet [19]. The whole optimized magnet modeled in OPERA-3D is shown in Fig. 3.

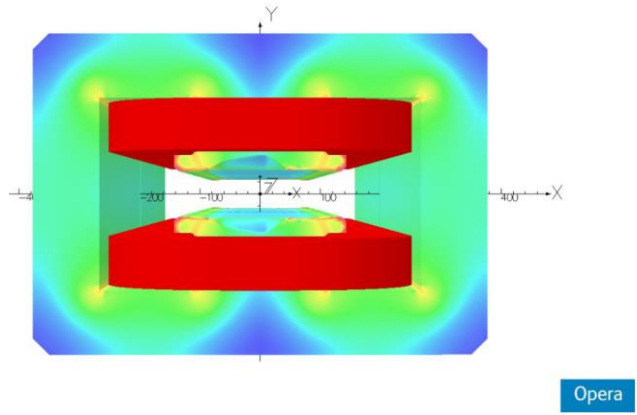

Fig. 3. LR-BB dipole magnet modeled in OPERA-3D (dimensions in mm)

The field homogeneity along the transverse axis in 2D plane at various longitudinal positions inside the iron core is investigated and depicted in Fig. 4. It can be seen that the transverse field uniformity in the GFR is within ±0.1% at all the longitudinal positions.

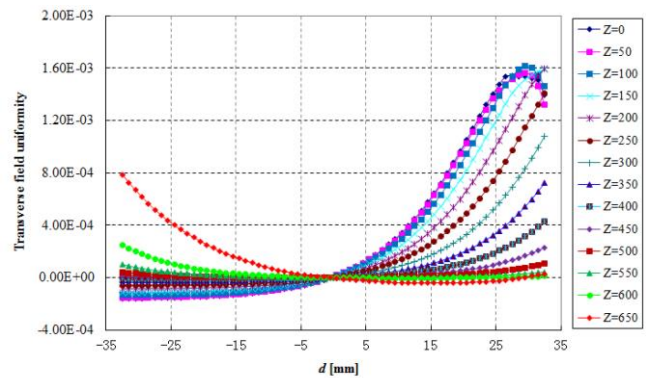

Fig. 4. Transverse field uniformity at various longitudinal positions

After the fabrication of the two LR-BB dipole magnets, magnetic field measurement was performed using a Hall probe system. The comparison of the measured and simulated integrated field distribution in the midplane ($y=0$) is presented in Fig. 5, where the central beam path corresponds to $d=0$.

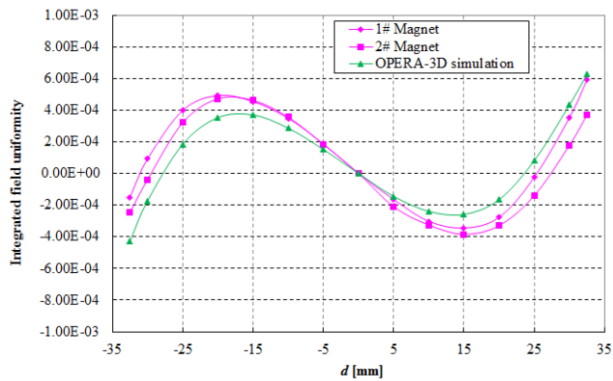

Fig. 5. The comparison of integrated field uniformity between the simulation and measurement results

It is shown that the measured integrated field distribution agrees well with the simulation result. The required integrated field uniformity of ±0.1% in the GFR is achieved. It should be noted that, when the field is integrated along straight lines as in the traditional method, the integrated field uniformity will be ±0.27% in the total GFR of 146.5 mm. So there is a large difference in the integrated field quality between the improved and traditional design methods for LR-BB dipole magnets.

The magnetic performance of the two LR-BB dipole magnets can well satisfy the field requirement, and these two magnets have been installed in the CSNS tunnel.

**3.2 Application to BEPCII 67B dipole magnet**

Since the beam trajectory is curved inside a dipole magnet, the integrated field uniformity obtained by integrating the field along curved beam paths in the improved design method is different from that in the traditional design method, and the former can be regarded as the actual field quality in a straight dipole magnet. The bending angle of CSNS LR-BB dipole magnets is as large as 25 degrees, and the beam sagitta is 81.5 mm. It is meaningful to study the field quality difference between the improved and traditional methods for an existing straight dipole magnet which has a smaller bending angle. So the improved design method is applied to 67B dipole magnet in the storage ring of the Beijing Electron Positron Collider Upgrade Project (BEPCII).

The BEPCII 67B dipole magnet is a laminated C-type straight dipole magnet with a pole gap of 67 mm, magnetic length of 1.4135 m and nominal central field of 0.6892 T (corresponding to the optimized beam energy of 1.89 GeV) [9, 20], which was developed using the tradition design method. Its pole end chamfer was determined by the field measurement result of prototype magnet in 2004. The fabrication and batch field measurement of forty 67B dipole magnets were finished in 2005, and they have been in operation since 2006.

The bending angle of BEPCII 67B dipole magnet is 8.8628 degrees, and the beam sagitta is 27.3 mm. The total width of horizontal GFR including the beam sagitta is 135.3 mm in the development of 67B dipole magnets, whereas the GFR required by the beam optics of 108 mm is used in the improved design method.

The whole 67B dipole magnet including the pole end chamfer is remodeled and the magnetic field analysis is performed using OPERA-3D. Then the field quality is investigated using the traditional and improved design methods respectively. Fig. 6 shows the whole 67B dipole magnet modeled in OPERA-3D.

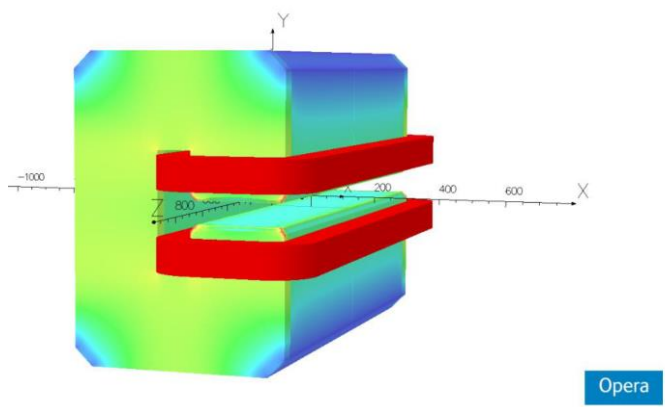

Fig. 6. 3D model of 67B dipole magnet (dimensions in mm)

The calculated integrated field distribution using the traditional design method is very close to the batch field measurement results of straight translating long coil system in 2005. The measured integrated field uniformity in the midplane of the batch 67B magnets using the traditional design method at the nominal excitation current is within $\pm 2\times 10^{-4}$ [20], whereas the calculated one is $\pm 1\times 10^{-4}$ in OPERA-3D.

Fig. 7 presents the comparison of the integrated field homogeneity in the two design methods, where the central beam path corresponds to $d$=0 in the improved design method and the mechanical centerline denotes $x$=0 in the traditional design method, respectively.

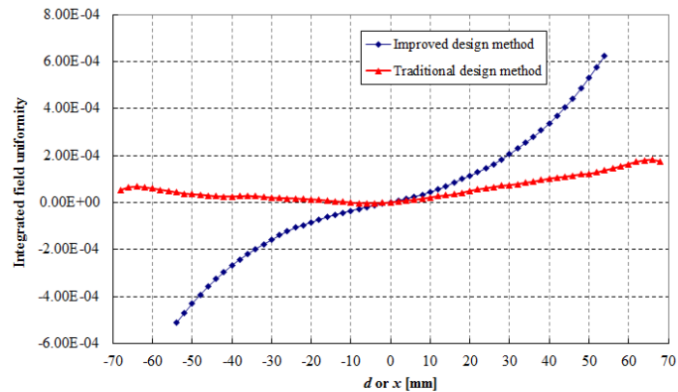

Fig. 7. Calculated integrated field uniformity of 67B dipole magnet in two methods

It is shown that there is a significant difference in the integrated field quality between the two methods for BEPCII 67B dipole magnet. The calculated integrated field uniformity is only $\pm 1\times 10^{-4}$ in the traditional design method, but it is as large as $\pm 6\times 10^{-4}$ in the improved design method. The latter one represents the accurate theoretical integrated field quality in the BEPCII 67B dipole magnet.

## 4 Conclusion

In the improved design method, the process in modeling and fabrication of the straight dipole magnet is the same to that in the traditional design method, but the field analysis and field measurement is similar to the ones in a curved dipole magnet.

The improved design method is applied successfully to develop the CSNS LR-BB straight dipole magnets, and is also applied to the existing straight 67B dipole magnet in BEPCII storage ring to illustrate the field quality difference between the improved and traditional design methods. The discrepancy in integrated field quality between the two methods cannot be neglected.

The integrated field quality obtained using the traditional design method is not accurate, and the one in the improved design method represents the actual field quality in a straight dipole magnet. Using the improved design method, the pole width, magnet dimension and weight of a straight dipole magnet can be reduced compared with the ones in the traditional design method. The improved design method can be applied to future straight dipole magnets.